\documentstyle[aps,twocolumn,prl,epsf]{revtex}

\begin{document}

\newcommand{\be}{\begin{equation}}
\newcommand{\ee}{\end{equation}}
\newcommand{\bn}{\begin{eqnarray}}
\newcommand{\en}{\end{eqnarray}}

\draft

\twocolumn[\hsize\textwidth\columnwidth\hsize\csname @twocolumnfalse\endcsname

\title{Dynamical correlations in half-metallic manganites}

\author{L. Craco$^1$, M. S. Laad$^2$ and E. M\"uller-Hartmann$^2$}

\address{${}^1$ 
Max Planck Institute for the Physics of Complex Systems, 
D-01187 Dresden, Germany 
\\  
${}^2$Institut fuer Theoretische Physik, Universitaet zu Koeln, Zuelpicher 
Strasse, D-50937 Koeln, Germany  
}
\date{\today}
\maketitle

\widetext

\begin{abstract}
Motivated by the recent optical and photoemission measurements on 
half-metallic ferromagnetic three-dimensional manganites, we combine a 
tight-binding fit of the one-particle bandstructure with the dynamical 
mean-field theory, which treats the dynamical orbital correlations and 
the combination of Jahn-Teller and doping-induced disorder on the same 
footing. We show how all of the above effects are necessary to obtain 
good semiquantitative agreement with experimental features. As applications, 
we show how modest external magnetic fields drive drastic changes in the 
optical spectrum, demonstrating the {\it colossal ac magnetoconductivity}. 
The photoemission lineshape contribution is evaluated, and good agreement 
with published experimental work is found.
\end{abstract}
     

]

\narrowtext

In view of its potentially large technological applications, the colossal 
magnetoresistance (CMR) has attracted renewed attention in the past few years.
The sensitivity of transport and optical response to small magnetic fields 
promises to open up new, attractive applications in optoelectronic 
devices~\cite{[1]}. A proper analysis of this problem requires combination of 
realistic material aspects (captured in one-electron bandstructure) with 
strong correlation effects, which are known to determine the physical 
response of transition metal compounds. Such approaches, recently been 
applied~\cite{[2]} to the study of photoemission spectra of materials 
believed to undergo filling-driven Mott transitions, have shown the 
importance of including proper bandstructure effects into a state-of-the-art 
correlation calculation. However, generically, one has to deal with the added 
complication of orbital degeneracy and associated Jahn-Teller distortions, 
which maybe static or dynamic~\cite{[3]}. This is hard in practice,
requiring one to deal with many bands-in this situation, a simpler approach 
to model the actual bandstructure, keeping the electronically active states 
intact, is an attractive option.  This is the main motivation of the 
tight-binding (TB) fits to the actual complex bandstructure, a strategy 
that has been used with success for the case of cuprates~\cite{[4]}. 

In this work, we use this strategy to study the field-dependent optical 
conductivity of well-doped manganites.  We concentrate on the well-doped, 
half-metallic ferromagnetic state in the cubic manganite 
$La_{0.7}Ba_{0.3}MnO_{3}$~\cite{[5]}. We combine the TB-fit with dynamical 
mean-field theory (DMFT) to treat the dynamical effects of local correlations 
with essential bandstructure features in a self-consistent way. We show how 
{\it all} the above effects are crucial to obtain a consistent picture of the 
field-dependent optical response and explicitly demonstrate the phenomenon 
of {\it colossal optical magnetoconductivity}. We also compute the 
photoemission lineshapes, pointing out its possible relevance to the 
study of the role of orbital correlations in the CMR materials.
  
We start by noticing that in the full linear augmented plane wave
(LAPW)~\cite{[6]} calculation, the 
majority $e_{g}$ bands, providing the conduction states, split from the 
$t_{2g}$ bands, but overlap with the O $p$ bands.  However, the two $e_{g}$ 
bands are clearly identified along the symmetry directions, making a mapping 
to an effective two-band model possible.  Constructing Mn-centered Wannier
functions with strong O $p\sigma$ character, only the $dd\sigma=t_{\sigma}$ 
and $dd\delta=t_{\pi}$ overlaps are nonzero.  Using Slater-Koster tables for 
the cubic perovskite structure, the $e_{g}$ bands are: 
$\epsilon_{\bf k}^{\pm}=\epsilon_{\bf k} \pm D_{\bf k}$, where 
$\epsilon_{\bf k}=(t_{\sigma}-t_{\delta})(c_{x}+c_{y}+c_{z})$,  

\bn
\nonumber
D_{\bf k}=-(t_{\sigma}-t_{\delta})
\sqrt{c_{x}^{2}+c_{y}^{2}+c_{z}^{2}-c_{x}c_{y}-c_{y}c_{z}-c_{z}c_{x}} \,
\en
and $c_{\alpha} \equiv cos(k_{\alpha}a)$. The full $e_{g}$ bandwidth is  
$2D=6(t_{\sigma}+t_{\delta})$.  To fix parameters above, notice that the 
dispersion along $\Gamma-X$ is equal to $4t_{\delta}$.  Since the overlap 
of the majority $e_{g}$ bands is omitted in the two-band fit, one uses the 
average value of the $\Gamma-X$ dispersion, giving $4t_{\delta}=-0.12$, or
$t_{\delta}=-0.03$ eV.  From $D=2.15$ eV, one has $t_{\sigma}=-0.69$ eV.
The Fermi level is set to zero by choosing $\epsilon=0.81$ eV, completing 
the simple TB fit to the actual LAPW bandstructure.
The corresponding total DOS~\cite{Weisse} is clearly different from the 
model DOS used in usual model hamiltonian treatments, and describes the 
doubly-degenerate $e_{g}$ bandstructure in the cubic perovskite geometry.
Starting from similar form for the bandstructure Wei$\ss$e 
{\it et. al.}~\cite{Weisse} concentrate on the polaronic physics 
while ignoring the undoubtedly strong local, orbital correlations in the FM 
phase~\cite{[5]}.  It is precisely our aim to show how a combination of 
strong correlations with Jahn-Teller distortion-induced orbital disorder 
and proper one-electron bandstructure is required for a qualitative (in our 
case, we found even semiquantitative) agreement with published experimental 
results.

With the single-particle dispersion relation from the TB-fit, the 
one-electron part of the hamiltonian is,

\bn
\nonumber
H_{0}=\sum_{\bf {k}\alpha}
\epsilon_{{\bf k}\alpha} c_{{\bf k}\alpha}^{\dag} c_{{\bf k}\alpha} \;,
\en
where $\alpha=a,b$ label the orbital indices for the doubly degenerate $e_{g}$
sector for the case of the cubic manganites.  The usual spin index is dropped
since one of the spin species is projected out of the problem in the 
double-exchange limit~\cite{[7]}. In this situation, the interaction part 
of the hamiltonian containing inter-orbital coulomb interaction and the 
Jahn-Teller coupling is given by~\cite{[8]},

\bn
\nonumber
H_{int} & = & U_{ab}\sum_{i}n_{ia}n_{ib} - g\sum_{i}
(Q_{i2}\tau_{i}^{z}+Q_{i3}\tau_{i}^{x}) \\ \nonumber
& + &  \frac{k}{2}\sum_{i}(Q_{i2}^{2}+Q_{i3}^{2}) \;.
\en
  
The $J_{H}\rightarrow\infty$ limit introduces a DE projection 
factor~\cite{[9]} into the hopping term, which depends on the nearest 
neighbor core-spin correlation function: 
$t_{ij} \rightarrow t_{ij} \sqrt{1+<{\bf S_{i}}.{\bf S_{j}}>/2S^{2}}$, 
and is temperature and magnetic-field dependent.  The one-particle part 
is now written as,

\bn
\nonumber
H_{0}=\sum_{<ij>\alpha\beta}t_{ij}^{\alpha\beta}(M)
(c_{i\alpha}^{\dag}c_{j\beta}+h.c) \;.
\en

The total hamiltonian of the system in the FM phase is thus $H=H_{0}+H_{int}$.
To simplify matters further, we make a local rotation in the two-dimensional
$Q_{2}-Q_{3}$ space, so that the local quantization axis is parallel to $z$.
One uses a local unitary transformation 
$U_{i}^{\dag}[{\bf Q_{i}}.{\bf \tau_{i}}]U_{i}=Q_{i}\tau_{i}^{z}$  
and simultaneously transforms the electronic operators as 
$a_{i\alpha}=U_{i}c_{i\alpha}$.  The hamiltonian then reads,

\bn
\nonumber
H &=& -\sum_{<ij>\sigma\sigma'}t_{ij}^{\sigma\sigma'}(M)U_{i}^{\dag}U_{j}
(a_{i\sigma}^{\dag}a_{j\sigma'}+h.c) + 
U_{ab}\sum_{i}n_{i\uparrow}n_{i\downarrow} \\ \nonumber
& - & g\sum_{i\sigma}Q_{i}\sigma n_{i\sigma}
\en
omitting the purely phononic part, and relabelling
$a_{i\uparrow}=a_{i}$ and $a_{i\downarrow}=b_{i}$.  With this, we have 
simplified the two-band model in the DE limit to a Hubbard-like model with 
a disordered ``magnetic field'' whose source is the disordered Jahn-Teller 
distortions which exist in the well-doped FM state at low $T$. 

A plausible parameter range for the problem at hand is $U_{ab}>D$ and 
$gQ\equiv v \le U_{ab}$, requiring one to treat the dynamical effects of 
strong local correlations and repeated scattering produced by moderately 
strong, local disorder on the same footing. This is achieved by using the 
DMFT as an approximation to our $3d$ problem. In $d=\infty$, the full lattice 
problem is mapped onto a single site problem, with the ``impurity'' embedded 
self-consistently in a dynamical bath.  In the multiband case, and without 
symmetry breaking (orbital ordering in our case), the  Green function and the 
purely local self-energy are functions of frequency only: 
$G_{nn'}(\omega)=G(\omega)\delta_{nn'}$ and 
$\Sigma_{nn'}(\omega)=\Sigma(\omega)\delta_{nn'}$.  In this situation, 
the DMFT selfconsistency condition becomes,

\bn
\nonumber
G_{mm'}(\omega,M)=\frac{1}{N} \sum_{\bf k} \frac{1}
{(\omega-\Sigma(\omega))\delta_{mm'}-(H_{0}^{TB}({\bf k},M))_{mm'}} \;.
\en
   
The remaining problem is to compute the single-particle self-energy, 
$\Sigma(\omega,M)$, in a situation where strong electronic correlations 
and disorder-induced strong scattering are simultaneously dominant, as above.  
We treat the strong orbital correlations at the local level using the 
iterated perturbation theory (IPT) away from half-filling, and generalized 
to finite temperature~\cite{[8]}, and the repeated scattering effects of the 
local Jahn-Teller (disordered) distortion by a proper combination of the 
IPT with the coherent-potential approximation (CPA), which solves the static 
disorder problem exactly in $d=\infty$~\cite{[10]}. In this procedure, the 
IPT self-energy is fed into the CPA local potential, which
now becomes $V_{i}=v_{i}-\Sigma_{int}(\omega)-\Sigma_{CPA}(\omega)$.  The IPT
local propagator is then used alongwith this in the disorder-averaged 
$T$--matrix (CPA) equation (but with 
$\Sigma(\omega)=\Sigma_{int}(\omega)+\Sigma_{CPA}(\omega)$): the 
local GF used in the CPA equation is                             

\bn
\nonumber
G(\omega,M)=\frac{1}{N} \sum_{{\bf k},\pm} \frac{1}{\omega-\Sigma_{int}
(\omega,M)-\Sigma_{CPA}(\omega,M)-\epsilon_{\bf k}^{\pm} } \;.
\en

Solution of the CPA equation $\langle T_{ii}[\Sigma(\omega)]\rangle = 0$,
gives the total self-energy corrected simultaneously for scattering caused 
by interactions (IPT) and disorder (CPA). This is fed back into the IPT 
subroutine that calculates a new self-energy, $\Sigma_{int}^{new}(\omega)$ 
and a new local GF, $G^{new}(\omega)$. These are fed back into the modified 
CPA routine, and the procedure is iterated to selfconsistency. The  
interacting density of states (DOS) is then obtained from the usual equation, 
$\rho(\omega,M)=-Im G(\omega,M)/\pi$. 

\begin{figure}[h]
\epsfxsize=3.5in
\epsffile{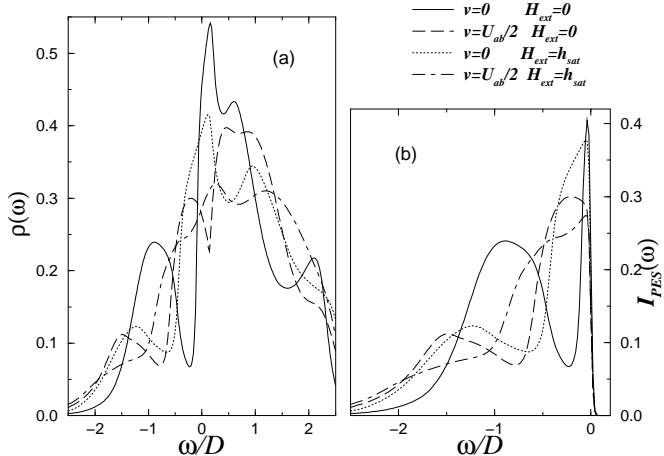}
\caption{(a) shows the spectral DOS for $n=0.6$ and $U_{ab}=2D$.
(b) shows the corresponding photoemission lineshape in the non-half-filled 
regime for $T=0.01D$.  The two values of the JT disorder and the 
external magnetic field are shown on the top of the figure.}
\label{fig1}
\end{figure}

We now present the results.  Below, we concentrate on a non-half-filled band
$n=0.6$ at low temperature regime $T=0.01D$.
At ``half-filling'', the true state is a Neel-ordered orbital antiferromagnet 
with additional correlated distortions coming from the JT coupling. To study 
the well-doped FM regime with ``melted'' orbital order, we consider only the 
para-orbital state. In Fig.~\ref{fig1}~(a), we show results for the local 
spectral function (DOS) for $U_{ab}=2.0D$.  For $v=0$, it shows features 
associated with the development of a correlated Fermi liquid metal associated 
with collective screening of ``orbital'' moments, in analogy with what 
happens in the usual $d=\infty$ Hubbard model.  However, we observe a rich 
structure of the DOS directly related to the use of the TB-bandstructure 
instead of idealized model input bandstructure used in model hamiltonian 
treatments.  

The effect of moderate local JT disorder is modeled by a binary disorder
distribution: $P(Q_{i})=x\delta(Q_{i})+(1-x)\delta(Q_{i}-Q)$, where $x$ is the
concentration of the $Mn^{4+}$ sites upon doping.  This is justified because
hole doping creates locally JT-inactive $Mn^{3+}$ sites, which are randomly 
distributed in the host lattice of JT active $Mn^{4+}$ ions. 
To make close contact with experiment~\cite{[17]} we choose $x=0.3$.
In Fig.~\ref{fig1}~(a) we show our results for the two values of $t_{ij}$, 
i.~e., for $t_{ij}=D$ and $t_{ij}=\sqrt{2}D$ and for $v=0,U_{ab}/2$. 
In particular, beyond a critical $v=v_{c}$, the spectrum is completely 
incoherent, due to strong (resonant) scattering off the disordered JT 
potentials.  The DOS is characterized by a low-energy pseudogap, resulting 
in an anomalous response.  

Next, we turn to the photoemission lineshape: 
$I_{PES}(\omega)=f(\omega-\mu)\rho(\omega)$.
With $v=U_{ab}/2$ and $t_{ij}=D$, we see only a small ``coherent'' 
weight at the Fermi level $(\mu)$, as shown in Fig.~\ref{fig1}~(b).  
The overall details of the calculated spectrum resemble the experimental 
result~\cite{[17]} quite well. An external magnetic field, $H_{ext}=h_{sat}$ 
results in increased coherence, as shown by the increased coherent component 
at $\omega=\mu$. The role of dynamical orbital correlations and JT 
distortions is again clear.  In the pure DE models, one expects a large 
quasiparticle contribution at low-$T$, and the DE model combined with 
disordered JT distortions yields a pseudogapped spectrum with no coherent 
contribution at low energy.  Within our approach, we have succeeded in 
obtaining the quasicoherent low-energy and the incoherent high-energy 
satellite features in semiquantitative agreement with experimental 
work~\cite{[17]}. Notice the additional shoulder structure in our 
calculations on the higher-energy side of the coherent feature. This is a 
direct consequence of the realistic features in the TB-fit spectral DOS and 
would be missed by model bandstructures. Finally, in actual practice, surface 
effects and possible inhomogeneities in the sample correspond to a larger 
$U_{ab}$, further enhancing the pseudogap feature in agreement with experiment.

The optical conductivity is computed from the Kubo formula. In $d=\infty$,
the vertex part does not enter the Bethe-Salpeter equation for the 
conductivity~\cite{[11]} so that,

\bn
\nonumber
\sigma_{xx}(\omega,M) & = & \sigma_{0}\sum_{\sigma}\int \rho_{0}
(\epsilon)d\epsilon \int d\omega' 
D^{2}(M)A_{M\sigma}(\epsilon,\omega')\\ \nonumber
&\times& A_{M\sigma}(\epsilon, \omega+\omega') 
\frac{f(\omega')-f(\omega'+\omega)}{\omega}
\en

\begin{figure}[htb]
\epsfxsize=3.5in
\epsffile{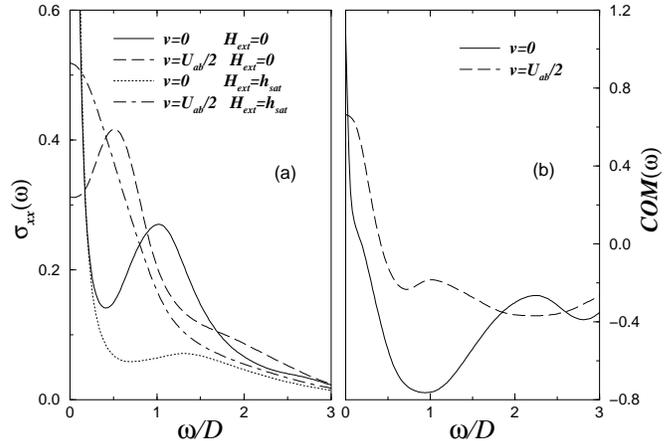}
\caption{(a) shows the optical conductivity for the same parameters as in 
Fig.~\ref{fig1}~(b).
(b) shows our results for the collossal optical magnetoconductivity 
for two values of the JT disorder.}
\label{fig2}
\end{figure}

The calculated $\sigma_{xx}(\omega)$ in $d=\infty$, clearly shows up the 
magnetic-field induced spectral weight transfer in the ``bad metal'' state 
above $T_{c}^{FM}$ for $x=0.3$.  In Fig.~\ref{fig2}~(a), we show 
$\sigma_{xx}(\omega,M)$ with the above parameters for
two values of $t_{ij}$, which correspond to $H_{ext}=0~(D(M)=D)$ and 
$H_{ext}=h_{sat}~(D(M)=\sqrt{2}D)$, as explained above. With 
$H_{ext}=0$ and $v=U_{ab}/2$, the Drude-like part in $\sigma_{xx}(\omega)$ at 
$T > T_{c}^{FM}$ is strongly suppressed.  This suppression is caused by
the combined effects of the decrease of $D(M)$ at high-$T$, the
$T$-dependent reduction of the spectral weight (due to the $T$-dependence 
of the quasicoherent collective Kondo peak in our Hubbard-like model in 
$d=\infty$) and strong scattering off disordered (local) JT distortions, 
which by itself would tend to open up a dip feature in the DOS.  
With $H_{ext}=h_{sat}$, we see a strong enhancement of the optical spectral 
weight in the low- and mid-infrared part of the spectrum. This is understood 
in terms of the physics of the $d=\infty$ Hubbard-like model, where 
$U_{ab}/D(M)$ controls the distribution of spectral weight; for large 
$U_{ab}/D(M)$, most of the weight is concentrated in the high-energy part, 
and a small fraction comprises the coherent part (in the para-orbital 
metallic state), in agreement with the fact that the resistivity magnitude 
places the material in the ``bad metal'' class. With increasing $D(M)$, 
high-energy spectral weight is transferred over energies on the scale of 
$U_{ab}$ to the quasicoherent part.  This transfer scales with the form of 
$D(M)$ as $D^{2}(M)=D^{2}(1+M^{2})/2$, completely consistent with experiment.

What is interesting is that at low energies, the fractional increase 
$\Delta\sigma_{xx}(\omega,M)/\sigma_{xx}(\omega,0)$ is a few hundred 
percent! (see Fig.~\ref{fig2}~(b)). We call this {\it colossal optical 
magnetoconductivity} (COM). Amazingly, exactly such a phenomenon has been 
observed by Boris {\it et al.}~\cite{[12]} for $La_{0.67}Sr_{0.33}MnO_{3}$.  
In practice, $H_{ext}$ is a few teslas, and, alongwith the fully 
spin-polarized nature of the ferromagnetic metallic state, this phenomenon 
makes for interesting field-dependent optical applications. We emphasize 
the crucial role of dynamical orbital correlations and JT effects here: the 
$T$ and $M$-dependent transfer of spectral weight has no analogy in a 
non-interacting system.   A similar calculation of $\sigma_{xx}(\omega)$ 
by Millis {\it et al.}~\cite{[13]} for the case $U_{ab}=0$ gives a 
completely incoherent optical response, in contradiction with experiment, 
and approaches based on the pure DE model in the cubic geometry of the 
perovskites~\cite{[14]} overestimate the quasicoherent part vis-a-vis 
experiment, and completely miss the mid-IR features~\cite{[15]}. 

In our picture, the extreme sensitivity of the ac response to a field of a 
few teslas arises via the change in the hopping, $t_{ij}^{\sigma\sigma'}(M)$, 
which transfers optical spectral weight over large energy scales of $O(D)$. 
The observation of Simpson {\it et al.}~\cite{[16]} also has a natural 
interpretation in terms of Hubbard model like physics in the orbital sector, 
as shown above. In Fig.~\ref{fig2}~(b), we also show how the COM is reduced 
by static JT disorder, demonstrating that the COM is related to the increased 
field-induced itinerance of the $e_{g}$ holes via 
$t_{ij}^{\sigma\sigma'}(M)$.     

In conclusion, we have shown how a combination of essential bandstructure
aspects of the cubic perovskite structure with a DMFT treatment of strong
orbital correlations in the $e_{g}$ sector in the DE limit describes the 
optical and photoemission spectra of the metallic CMR manganites in a 
semiquantitative way. Additional interesting applications of our modelling 
to investigate field-dependent magneto-optical response across the 
para-ferro transition are being studied, and will be reported elsewhere.
Our treatment can be extended to include other broken symmetries in the spin 
and orbital sectors, and should be generally applicable, with suitable
modifications, to other half-metallic TM-oxide ferromagnets.

MSL acknowledges the financial support of the SfB341 of the German DPG.
LC was partially supported by the Funda\c c\~ao de Amparo \`a Pesquisa 
do Estado de S\~ao Paulo (FAPESP) 
and by the Max-Planck-Institute f\"ur Physic komplexer Systeme. 
LC wishes to acknowledge the kind hospitality 
of the Instituto de F\'{\i}sica Gleb Wataghin~(UNICAMP) 
and the Institute for Solid State and Materials Research Dresden.


\begin{references}

\bibitem{[1]} M. Imada, A. Fujimori and Y. Tokura, Rev. Mod. Phys. 
{\bf 70}, 1039 (1998).

\bibitem{[2]} see, for example, A. Georges, G. Kotliar, W. Krauth and M. 
Rozenberg, Rev. Mod. Phys. {\bf 68}, 13 (1996).  
Also, K. Held, G. Keller, V. Eyert, D. Vollhardt, and V. I. Anisimov,
Phys. Rev. Lett. {\bf 86}, 5345 (2001) for application to the 
insulator-metal transition in $LaTiO_{3}$ and doped $V_{2}O_{3}$.

\bibitem{[3]} for a discussion relevant to the titanates, see 
M. Imada, preprint cond-mat/0105135;
the relevance of orbital correlations to the physics of manganites is dealt
with by many authors, for example, S. Yunoki, A. Moreo and E. Dagotto, in
{\it Physics of Manganites}, eds. A. Kaplan and S Mahanti, Plenum Press 
(1998).

\bibitem{[4]} Such a strategy for the cuprates has been widely used for 
the cuprates by E. Pavarini, I. Dasgupta, T. Saha-Dasgupta, O. Jepsen, 
O. K. Andersen, preprint cond-mat/0012051.

\bibitem{[5]} In this paper, we consider only the ferromagnetic 
half-metallic state.
An implicit assumption is that AF-orbital order that exists in the undoped
parent compounds is melted upon hole-doping, but local orbital correlations 
are still dominant even in the well-doped FM state.  See, for example,
K. Held and D. Vollhardt, Phys. Rev. Lett. {\bf 84}, 5168 (2000); 
also, M. S. Laad, L. Craco and E. M\"uller-Hartmann, 
Phys. Rev. B {\bf 63}, 214419 (2001).

\bibitem{[6]} see, for example, W. Pickett and D. Singh, in 
{\it Physics of Manganites}, eds. A. Kaplan and S Mahanti, 
Plenum Press (1998).

\bibitem{Weisse} A. Wei$\ss$e, J. Loos, and H. Fehske, Phys. Rev. B {\bf 64},
104413 (2001).

\bibitem{[7]} N. Furukawa, J. Phys. Soc. Jpn. {\bf 63}, 3214 (1994).

\bibitem{[9]} P. W. Anderson and H. Hasegawa, Phys. Rev. {\bf 100}, 
675 (1955). 

\bibitem{[8]} M. S. Laad, L. Craco and E. M\"uller-Hartmann, 
preprint cond-mat/0101335, submitted to Phys. Rev. B.

\bibitem{[10]} L. Craco and K. Kang, Phys. Rev. B {\bf 59} 12244 (1999);
M. S. Laad, L. Craco and E. M\"uller-Hartmann, 
cond-mat/9911378, accepted for publication in Phys. Rev. B.

\bibitem{[17]} T. Saitoh, D. S. Dessau, Y. Moritomo, T. Kimura, Y. Tokura and
N. Hamada, Phys. Rev. B {\bf 62}, 1039 (2000).

\bibitem{[11]} A. Khurana, Phys. Rev. Lett. {\bf 64}, 1990 (1990).

\bibitem{[12]} A. V. Boris, N. N. Kovaleva, A. V. Bazhenov, 
P. J. M. van Bentum, Th. Rasing, S-W. Cheong, A. V. Samoilov, and N.-C. Yeh, 
Phys. Rev. B {\bf 59}, R697 (1999). 

\bibitem{[13]} A. J. Millis, H. M\"uller, and B. I. Shraiman,
 Phys. Rev. B {\bf 54}, 5405 (1996).

\bibitem{[14]} P. E. Brito and H. Shiba, Phys. Rev. B {\bf 57}, 1539 (1998), 
see also,  H. Shiba, R. Shiina, and A. Takahashi, J. Phys. Soc. Jpn. 
{\bf 66}, 941 (1997). 

\bibitem{[15]} the authors of~\cite{[14]} propose that the mid-IR features 
originate from interband transitions in the $e_{g}$ sector.  While we 
believe this to be the case, the results 
of~\cite{[14]} lead to inconsistencies with experiment in that (i) the 
Drude part is overestimated, and (ii) the optical mass enhancement is in 
conflict with that extracted from the low-$T$ specific heat. 

\bibitem{[16]} J. R. Simpson, H. D. Drew, V. N. Smolyaninova, R. L. Greene, 
M. C. Robson, Amlan Biswas, and M. Rajeswari, Phys. Rev. B {\bf 60}, 
R16263 (1999).
\end{references}
\end{document}